# Interaction Between Two Ring Shaped YBaCuO Superconductors with currents $i_a$ and $i_b$

G. Briones Galán, G. Domarco, L. Romaní

Departamento de Física Aplicada, Universidad de Vigo , As Lagoas s/n,32004 Ourense, Spain

**Abstract**

Two rings with currents ia and ib become magnets that can either attract or repel each other. If mechanical work is extracted from this ring system, either by bringing closer the rings during an attraction or by separating them as the result of a repulsion, the energy of the rings, the only existing source of energy in the system, decreases: $E = \tfrac{1}{2} L I_a^2 + \tfrac{1}{2} L I_b^2$

If an energy either of attraction or repulsion is communicated to the system, (working against the forces) its energy increases, i.e. the currents in the rings increase. This happens because the currents existing in the rings are free, i.e. not linked to any device which imposes a current (power supply, constant current).

Similarly, the effect of electromagnetic induction can increase or decrease the currents with the corresponding mechanical work linked to these developments. Not only does the current in the rings decrease as they get closer during the attraction, but it also can be cancelled and, if the effect continues, it may even change its direction.

An application that already exists is the levitation train. It is attracted by the rail and held on it, but if it is very close to this due to the weight of the goods a strong repulsion (levitation) occurs. Another application would be the accumulation of energy in these rings, where they can spend large amounts of current, not only because of its high critical currents but also by the fact that the magnetic flux remains constant, cannot increase. Another advantage in this storage system is its complete reversibility in the exchange of electromagnetic energy and mechanical work.

In the energy storage system, the greater the current flowing through the ring, the greater its energy is ($\tfrac{1}{2} L I^2$). For the storage of important amounts of energy, a high current (I) is required. In order to attain this, there has been made a study of the ideal conditions (whether of attraction or repulsion) to store as much energy as possible. The higher the difference between initials currents, the higher difference between the magnetic forces of attraction and repulsion is. The mechanical work done in attraction (during an approach) reduces the magnetic energy into the ring and increases it in repulsion. The curves $F(x)$ for attraction and repulsion are not symmetric. The experiences carried out so far have proved the feasibility of a scheme based on superconducting ceramic rings, which would allow the reversible conversion of mechanical work into electromagnetic energy to be stored in the rings as superconducting permanents currents.

## 1. Introduction

The samples are made by the crystallization process "Top seeding melt grown". The seed is a Sm Ba$_2$ Cu$_3$ O$_7$ introduced into the fluid of Y Ba$_2$ Cu$_3$ O$_7$, whose melting point is higher than the YBCO and thus remains solid crystal driving the crystal growth during the slow cooling of YBCO. In the crystal lattice the green phase of YBCO and other types of doping are injected. During crystallization the formation of lattice defects (pinning) is favored to increase the amount of current. At the same time this method allows us that all the magnetic flux from the field is trapped by this material and behaves as a pure electrical conductor without Meissner effect. This simplifies the calculations of distribution of the permanent current to flow through the ring. Here the only property that has the superconducting ring is their zero electrical resistance

In this work have been studied interactions between superconducting rings made of YBaCuO, in order to explore the feasibility of a system for storing energy coming from mechanical work. This mechanical work is converted into electromagnetic energy in the form of permanent currents in the rings. The process is reversed to return reversibly again electromagnetic energy as mechanical work by the forces of attraction (or repulsion) between the loaded rings.

For this experiment confronted two of these superconducting rings and measured the attractive and repulsive force between them as function of its separation x.

The Faraday-Lenz Law (from now FLL) explains how a magnetic flux that goes through a closed current circuit acts over it. That is we will use to explain how bulk, ring shaped, superconductors interacts with each other when they have currents circulating through them.

Attending to the FLL, the flux passing through a ring can be written as:

$$\frac{\Delta \varphi}{\Delta t} = -V = -RI \qquad (1)$$

As we are talking about superconductors, we assume that $R=0$. Therefore we have,

$$\frac{\Delta \varphi}{\Delta t} = 0 \qquad (2)$$

This implies that $\varphi(t)$ = constant, the magnetic flux remains constant in this case. Consider a bulk YBaCuO ring called 'a' and another called 'b' which can move only vertically over a line passing through their center of symmetry as seen in figure1.

At the beginning we have two bulk superconducting rings 'a' and 'b' with an initial flux $\varphi_a$ and $\varphi_b$, $i_a$ and $i_b$ are the initials currents lower than critical currents.

Initial flux in a: $\varphi_i = L_a i_a$     Initial flux in b: $\varphi_i = L_b i_b$ (3)

If we approach sample 'b' to sample 'a' the magnetic fields will interact in attractive or repulsive way depending the faces we confront.

$\Phi = \varphi_a + \varphi_b$   attraction          $\Phi = \varphi_a - \varphi_b$   repulsion (4)

Where $\varphi_a$ and $\varphi_b$ are the magnetic flux for 'a' and 'b' respectively.

The magnetic flux in 'a' will be affected by 'b' and the flux in 'b' will be affected by 'a' too. This means that the final flux in 'a' and 'b' will be:

*Attraction* (5)
$L_a i_a = L_a I_a + M I_b$ = constant
$L_b i_b = L_b I_b + M I_a$ = constant

*Repulsion* (6)

$$L_a i_a = L_a I_a - M I_b = \text{constant}$$
$$L_b i_b = L_b I_b - M I_a = \text{constant}$$

Where $L_a I_a$ are the final flux crossing 'a' and $M I_b$ is the flux from 'b' which affects 'a'. In the same way, $L_b I_b$ is the final flux crossing 'b' and $M I_a$ is the flux from 'a' which affects 'b'. $M$ is the mutual inductance between 'a' and 'b'. From equations. (5) and (6), we can see that, in attraction, the final current is less than the initial current ($I_a < i_a$ and $I_b < i_b$) in both rings and the opposite in repulsion where the final current is higher than the initial one in both rings ($I_a > i_a$ and $I_b > i_b$), but never higher than critical currents. We can see now how the current varies with M in figure 2. Moreover, at a distance far enough, the interaction between the rings vanishes. Therefore, the mutual inductance M will be zero and the currents will remain constant in attraction and repulsion as we can see in figures

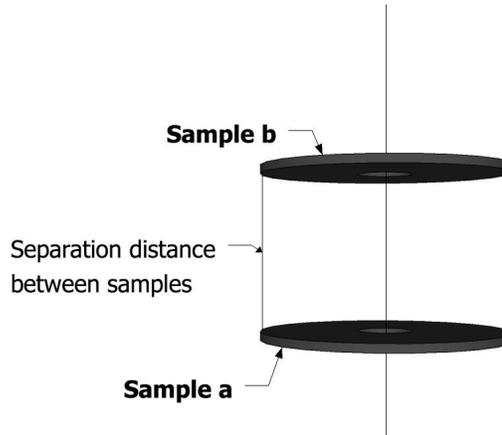

**Figure 1.** During the experiment; all samples will be confronted as shown in this figure. All the samples had a positive magnetic field on one face and a negative one on the other. This means that to change attraction to repulsion mode we only have to turn over the upper sample into the appropriate way.

From equations (5) and (6) we can obtain the final current for each ring.

*Attraction*

$$I_a = \frac{L_b(L_a i_a - M i_b)}{L_a L_b - M^2}$$

$$I_b = \frac{L_a(L_b i_b - M i_a)}{L_a L_b - M^2}$$

(7)

*Repulsion*

$$I_a = \frac{L_b(L_a i_a + M i_b)}{L_a L_b - M^2}$$

$$I_b = \frac{L_a(L_b i_b + M i_a)}{L_a L_b - M^2}$$

(8)

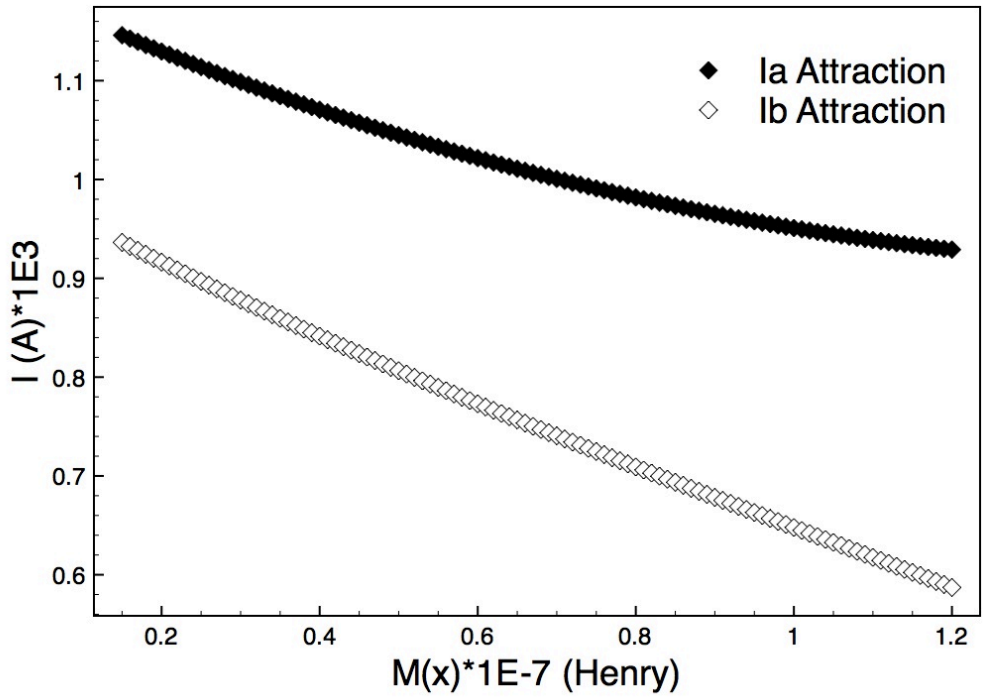

*(a) Theoretical curve when the magnetic force is attractive.*

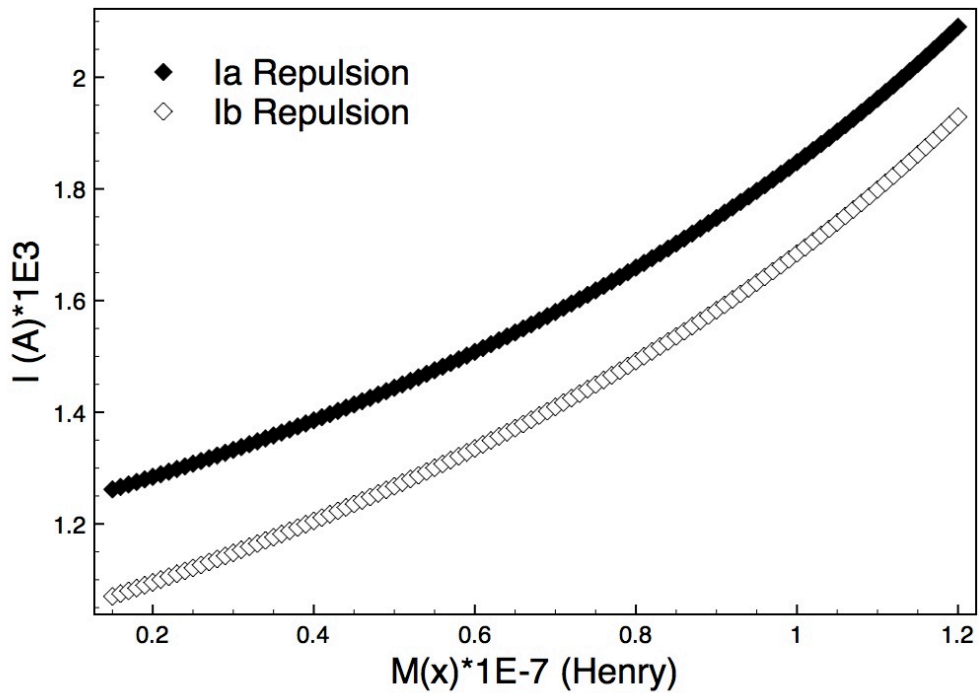

*(b) Theoretical curve when the magnetic force is repulsive.*

**FIG. 2. These figures represents the final values of I obtained from equations (7) and (8) starting from initial currents $i_a$ and $i_b$. We can see that M increases while distance between samples is getting closer.**

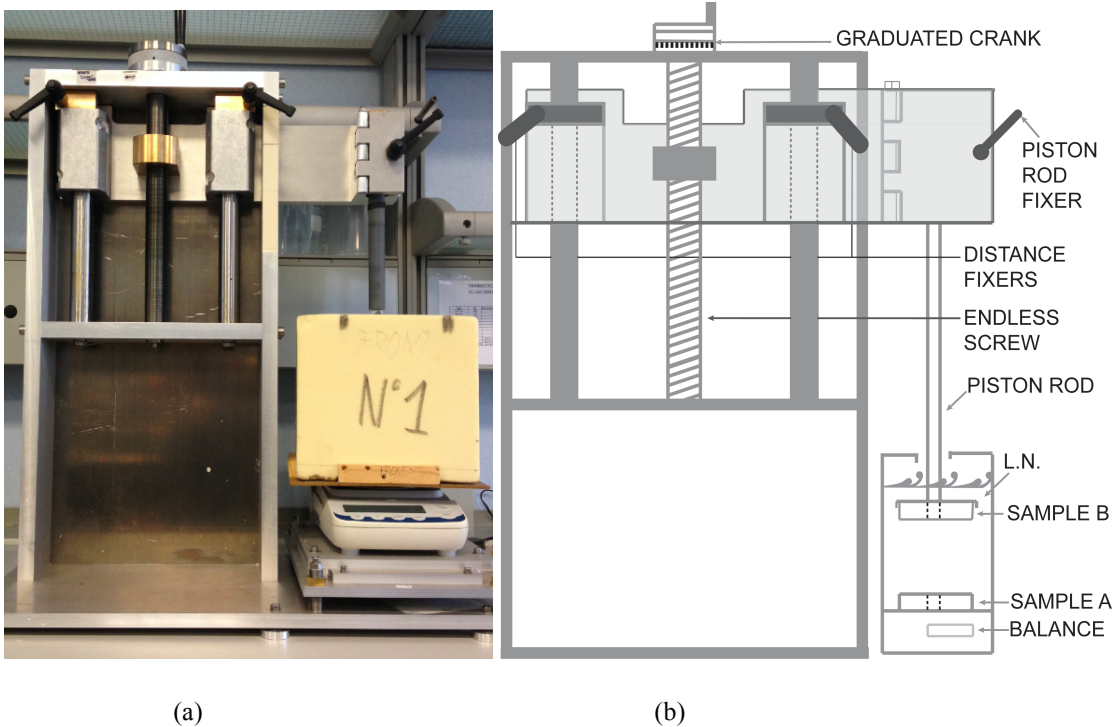

(a)  (b)

**FIG. 3.** This is the measurement machine (a) and the set up scheme (b). The block marked as N°1 is the cryostat placed over the balance and the piston rod is submerged into the liquid nitrogen. It could be seen too the crank and endless screw. Underneath the balance there is a leveling system to align the centre of the hollow with the sample fixed at the bottom and the sample at the end of the piston rod.

### 2. Experimental

The aim of the experiment is to measure the repulsion and attraction forces between two ring shaped superconductors by putting one above the other.

Samples were induced previously at constant field cooling with closed core [1]. We used rings with inner diameter of 10mm, outer diameter of 30mm and 3mm of thickness.

To measure the interaction forces between the induced rings we used a balance [2], [3], [5], [6], [7], [8]. Over the balance we place a polyurethane cryostat in whose bottom we fixed one of those two superconducting rings.

In every moment we kept submerged both rings in liquid nitrogen to ensure the superconducting state and avoid current losses. The sample at the bottom of the cryostat was fixed deep inside and does not move at all. The other one was attached to the end of a piston rod which moves from top to bottom over the sample, as shown in figure 3.

Maximum and minimum distance was scaled over the measure machine counting the relation between distances and rotating angle in a crank. This crank leads an endless screw which moves the piston rod up and down. Figure 3 shows a scheme of the procedure of measurement and a picture of the machine

#### 2.1. Measurement Process.

We measure the weight in the balance each time we approach the sample at the end of the piston to the bottom of the cryostat in which center were placed the fixed sample. The sample at the end of the piston rod approaches to the minimum distance of 10.5*mm* taking measurement from the balance each 0.5mm. Once the samples were separated at the minimum distance, the piston was moved up to the maximum separation distance of 30.5*mm*. Each time we measured at a fixed step and at regular time periods in order to replicate easily each time we measure.

In this context the forces which we consider involved are,

Cryostat weight ($F_C$)
Liquid nitrogen weight ($F_{LN}$)
Arquimedes push ($F_{Arq}$)
Magnetic force ($F_{mag}$)

The weight registered in the balance is the sum of all these effects. Therefore, we need to isolate the magnetic force from the other contributions.

$$F = F_C + F_{LN} + F_{Arq} + F_{mag} \tag{9}$$

Instead of this we isolated the other contributions from the magnetic forces making measurements with only the sample at the bottom of the cryostat and at the end of the piston we placed a nylon disc to avoid any magnetic interaction. Here we had all contributions except the contribution due to the magnetic force. We have then,

$$F_0 = F_C + F_{LN} + F_{Arq} \tag{10}$$

As we made the measurement process ensuring repeatability, the only we had to do was,

$$F - F_0 = F_{mag} \tag{11}$$

Doing this for all measurements we obtain the magnetic force $F_{mag}$.

*TABLE I. Technical characteristics of the superconducting rings used in this work. All samples were made by top seeding melt grown with Sm seed. Dimensions are 10 mm inner diameter, 30 mm outer diameter and 3 mm thickness. The i is the initial induced current measured by a Hall probe placed in the bottom of the ring's axis. According to provider, the critical current for these samples are about 10 kA.*

| Sample | P1 | P2 | P3 | P4 | P5 |
|---|---|---|---|---|---|
| i(A) | 5551 | 238 | 1586 | 872 | 634 |

In order to verify experimentally the decrease in *I* during the attraction and the increase in *I* during the repulsion, we use the experimental measurement of the force versus distance F(x); the force can be expressed by the equation (12), where $f(x) \geq x^2$ (it is a function increasing with x).

$$F(x) = Cte \frac{I_a I_b}{f(x)} \tag{12}$$

The absolute value F(x) increases, as x decreases, but at the same time the value of $I_a$ ($I_b$) decreases during the attraction and increases during the repulsion.
In an attraction, as the absolute value of F(x) increases, due to the decrease in x, there decreases the current *I* as well (see equation (7)); if the current is annulled, then F(x) is annulled, but if x keeps on decreasing, then F(x) eventually changes its sign. But the same does not happen to the absolute value F(x) during repulsion, since, as x decreases, the current *I* increases (see equation (8)) and never gets annulled. Both curves F(x), one under attraction, one under repulsion, are not symmetrical.
To highlight the decrease in current during the attraction, see Ec.7, choose a large initial current $i_a$ for the final current $I_b$ a change of sign appears. $i_a$ = 5551 A, $i_b$ = 238 A.

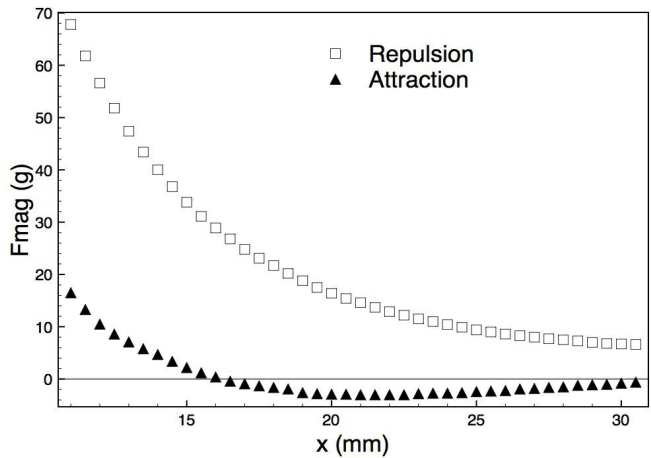

**FIG. 4.** Samples P1 and P2. During the repulsion the values of the current increases, as M increases (equation (8)), x decreases. The product Ia Ib increases continuously (equation (8)). Conversely the denominator f(x) decreases, but it always keeps its positive values. The value of the force increases as a result of both numerator and denominator. But during attraction the I's decreases as M increases (equation (7)). They even can be annulled and change its sign. See figures 5(a) and 5(b)

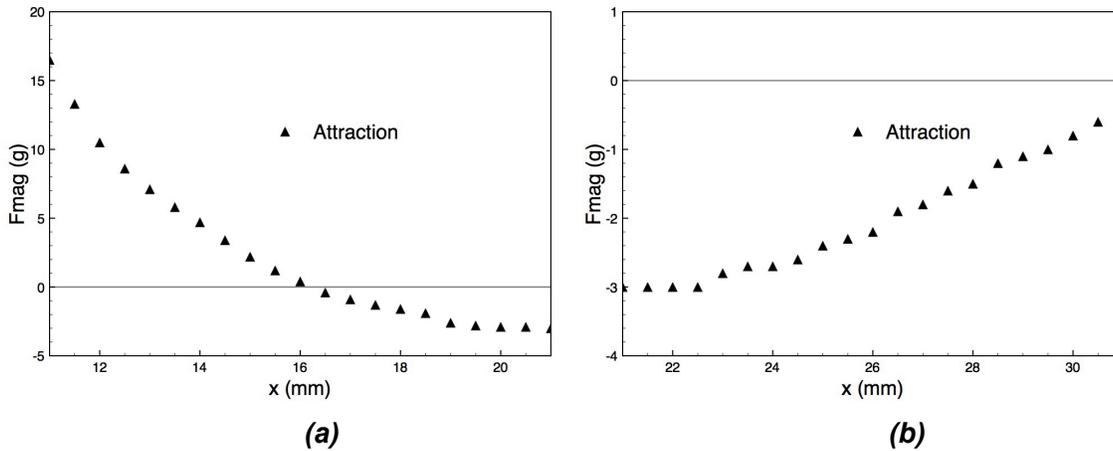

(a)                            (b)

**FIG. 5.** Attraction between samples P1 and P2. Here we can see how the sign changes as explained above.

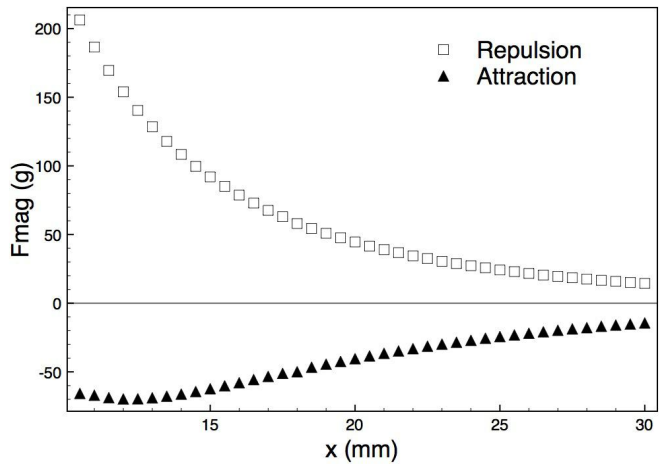

*FIG. 6. Samples P1 and P3. Attraction and repulsion curves.*

Now, in Figures. 7 and 8, the sign of the attracting force is changed in order to compare the values during attraction with the values during repulsion. In this way we can compare the mechanical work directly on the surface of the graphic. As a result, the values of the attracting forces prove to be always shorter than the repulsion forces. All of which confirms the fact that the currents under repulsion increase, while those under attraction decrease. If the samples are moved far enough from each other, so that the mutual induction M tends to zero, then the values under attraction and under repulsion tends to a single curve. That means, if the sign under attraction had not been changed, the curves would be symmetrical on this particular stretch. See figure 6 for x between 20 and 30 mm.

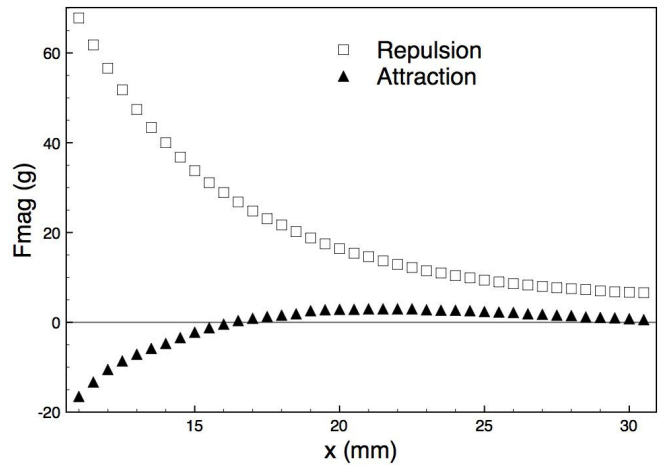

*FIG. 7. Samples P1 and P2. Here we change the sign in attraction of Fig.4 to obtain a better comparison between both curves.*

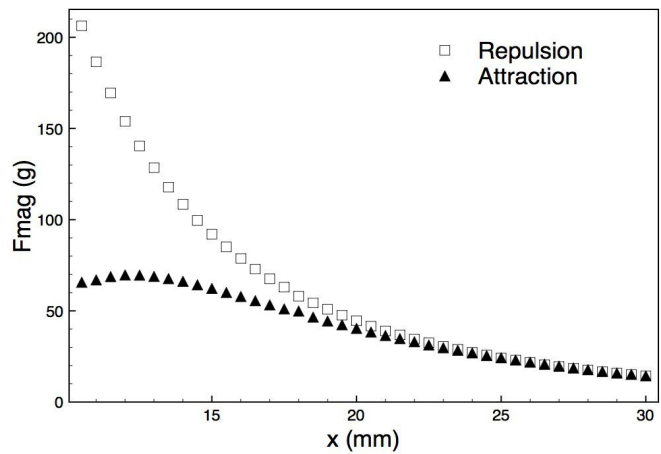

*FIG. 8. Samples P1 and P3. Here we change the sign in attraction of Fig.6 to obtain a better comparison between both curves*

## 2.2 Inertia in induction processes

To show the inertia that has an electromagnetic induction between two samples two measurements are made, a measure consists of a separation between the two samples and the other an approximation between them.
We started approaching one ring to another, and then distancing.
It may be a delay in establishing the final current

Mechanics                                                     Induction

$$F = m\,dv/dt \qquad\qquad\qquad V = L\,dI/dt$$

**On the circumference of a ring, $E = V / (2\pi R)$. For a charge q that is on the circumference,**

$$F = q\,E, \quad F = q\,V / (2\pi R).$$

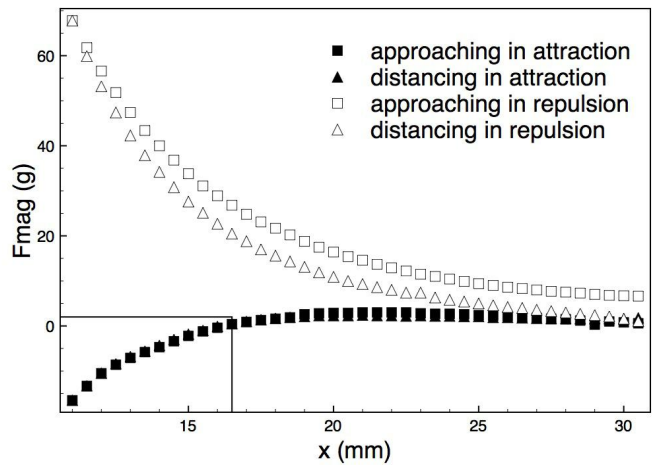

**FIG. 9.** Attraction and repulsion forces vs. distance between samples P1 ($i_a$ = 5551A) and P2 ($i_b$ = 238A). Forces are given in grams.

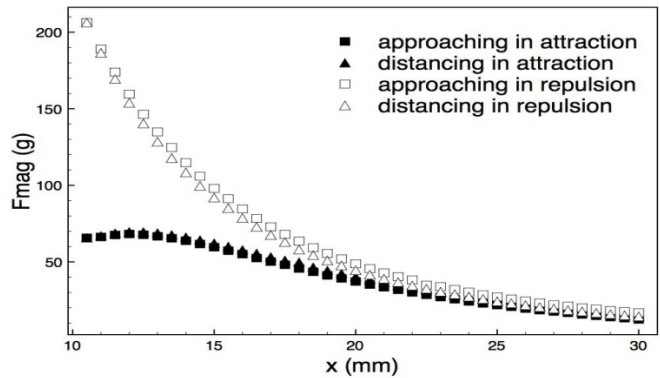

**FIG. 10.** Attraction and repulsion forces vs. distance between samples P1 ($i_a$ = 5551A) and P3 ($i_b$ = 1586A). Forces are given in grams

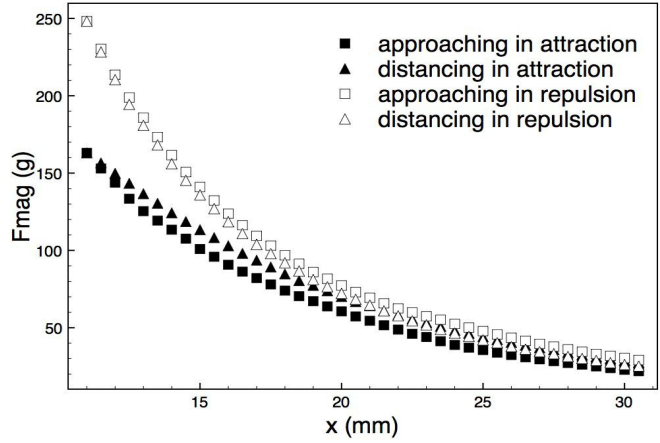

**FIG. 11.** Attraction and repulsion forces vs. distance between samples P4 ($i_a$ = 872A) and P5 ($i_b$ = 634A). Forces are given in grams

### 3. Results

Measurements show that magnetic forces $F_{mag}$ are proportional to the product of the induced current in both samples ($I_a \cdot I_b$), and it decreases as separation distance increases (12). All samples which we used in the experiment are included in table1:

We made measurements pairing P1-P2, P1-P3 and P4-P5. Current in P1 is 23 times higher than in P2 and 3.5 times higher than in P3. The induced current in P4 is about 1.37 times higher than in P5. The results can be seen in Figures 9, 10 and 11

By measuring the magnetic interaction forces with the distance we found that the induced current decreases in both samples during the magnetic attraction and increases during the magnetic repulsion

Charge carriers in superconductors are not affected by thermal agitation. Then, they are able to move along the crystalline net without interacting with any ion pinned in the net and without any resistance to its movement. Such charge carriers act like they were independent and moves from a nuclei to another one without getting disturbance by any of them. The other particles jointed to the nuclei take part in thermal agitation, even the nuclei.

In figure 9, $F_{mag}$ in repulsion and $F_{mag}$ in attraction diverge. At a distance of 16.2 mm, here $M_0 = L_b / 23$, $I_b = 0$ and $F_{mag} = 0$. If $M > M_0$, $I_b$ changes to negative and the forces are exerted in opposite direction (see Equation 7).

The difference between repulsion forces $F_R$ and attraction forces $F_A$ are,

$$F_R - F_A = \frac{2 L_a L_b M (L_a i_a^2 + L_b i_b^2)}{\left(L_a L_b - M^2\right)^2 f(x)} \tag{13}$$

Where $f(x) \geq x^2$. This difference is always positive and increases as increases M (decreases distance) as we saw in equations (7), (8) and (13). If $M = 0$, the difference is 0.

As the difference between initial currents $i_a$ and $i_b$ are too high in figure 9, the current decrement over $I_b$ during attraction is very strong and not only becomes zero but flows in opposite direction. In the same way we can see this behavior in figures 10 and 11 where the relation between $I_a$ and $I_b$ are not as high as in figure 9.

In a uniform field ($\partial H / \partial x = 0$) the force is zero [4]. In a material whose magnetic moment is M(A / m) there is only one force if there is a gradient near M ($\partial H / \partial x \neq 0$)

That is,
$$F = M \frac{\partial H}{\partial x} \left(\frac{Nw}{m^3}\right) \tag{14}$$

The work made by these magnetic forces is,

$$Fdx = VMdH \ (m^3A/mTesla) = Id\varphi \ (Joules) \tag{15}$$

## 4- Application to the levitation train.

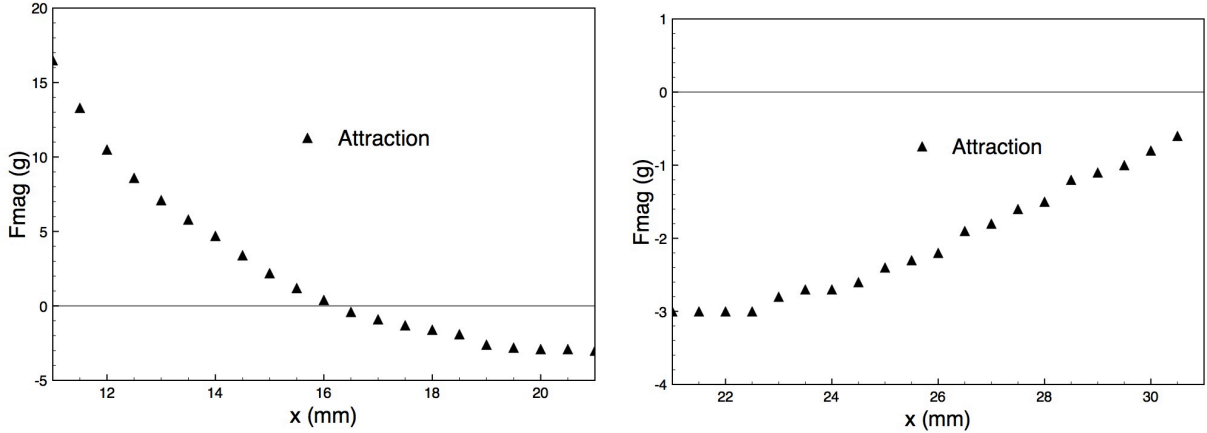

***FIG. 5 bis. Attraction between samples P1 and P2. Here we can see how the sign changes as explained above.***

The attraction and repulsion between two rings with free currents is not quite the same as the attraction and repulsion between two magnets. While in magnets the magnetic moments are fixed, in ferromagnetic materials, the domains may experience either an orientation or a disorientation. In magnets the orientation of the domains is fixed and has got a single direction, but in ferromagnetic materials the domains are generally disorientated.

In the presence of a magnetic field, the flux tubes of the field generate vortices in a superconductor. If the superconductor is massive, there is nothing but vortices; if the superconductor has a hole in its center, there may be currents too, provided that Ampere's circuital law can be established around the hole, $\oint \frac{B}{\mu_0} dl = I$. In a ferromagnetic material, the magnetic field orientates the domains in its direction. As the field increases, both the vortices and the domains come closer and closer to its direction. Between the source of the field and the sample there occurs an attraction, that is, both of them show a resistance to separating.

If the superconductor is separated from the magnetic field, the currents increase (that is, the rotation of the vortices increases in speed), and, vice-versa, if the superconductor comes close to the magnetic field, the rotation of the vortices decreases in speed; if it continues to approach the field, (that is, if the field continues to increase at the level of the superconductor), this decrease eventually brings the rotation to a full stop and reverses its direction, so that the former attraction between the rings becomes now a repulsion (Faraday- Lenz law).

In the presence of a magnetic field, the superconductor behaves as a ferromagnetic material, except that, instead of being there any rigid domains, in this case, there are vortices. The charge carriers are free and must satisfy the condition R = 0 (that is, constant flux). The interaction between a magnetic field and a superconductor applicable to the levitation train is the following:

Starting from a fixed point of magnetization in the superconductor, that is, a point on the rail, any separation from the rail causes an attraction between them, while approaching the rail causes a repulsion, that is, the train never touches the rail, but, since experiencing, at the same time, an attraction to it, remains over it. In the figure *5b* it can be seen that, despite the approaching between the rings, the current experiences such a slowdown that the force decreases clearly between them and reverses its direction.

## 5- Application to the storage of energy

In all experiments of interaction between two rings there can be observed the tendency to an exponential increase in the forces, as the rings approach each other, during the repulsion.
Let us see what happens theoretically, Equations. (7) and (8). Given two equal rings with the same initial currents: $L_a = L_b = L$ y $i_a = i_b = i$.

(Attraction) Eq.(7) $\quad I_a = I_b = I = \dfrac{Li\,(L-M)}{L^2-M^2} = \dfrac{Li}{L+M}$, $\quad$ if $\quad M \to L, \quad\quad I \to \dfrac{i}{2}$

(Repulsion) Eq. (8) $\quad I_a = I_b = I = \dfrac{Li\,(L+M)}{L^2-M^2} = \dfrac{Li}{L-M}$, $\quad$ if $\quad M \to L, \quad\quad I \to \infty$

While, during the attraction, both currents and forces tend to a constant finite value (i/2), during the repulsion, both of them increase as far as reaching very high values.
In conclusion, the repulsion has to be used, in order to store bigger amounts of energy.

## 6. Forthcoming research ensuing from these experiments

In a superconducting ring it may be observed both a slowdown and a change of direction of the current, but in a ferromagnetic material where the movements of the charge carriers behave like superconductors, currents come from ferromagnetic domains. But what happens when the movement of the currents comes to a stop or slow down?
 Pierre Weiss has supposed that the domains are magnetized to saturation; they are orientated so that the magnetic field is cancelled on the outside. Their effects may only be observed, if a magnetic field is capable of orientating the domains in a single direction. For this saturated magnetization, the charge carriers, that have got a mass as well, are supposed to exchange their energy with the inner energy of the medium until they reach an equilibrium with this medium, that is, the same temperature. But what happens when the energy of the domains decreases as a result of a mechanical work that leaves its system?
 In all possible oscillations of energy that our sample may undergo, there is only one source that can either absorb or loose energy: the inner energy of the medium where the sample is. These possible developments of ferromagnetic domains are now being researched by our team..

## 7. Conclusion

Circulating currents into the samples could be braked (or increased) by an inductive e.m.f. which acts over the charge carriers. This takes place because the mutual induction M acts as a vehicle for the induction between the samples. The attractive (or repulsive) mechanic work Fdx represents the energy lost (or gained) by charge carriers which are present in both rings, where the decreasing (or increasing) energy are determined by $Id\varphi$. The results attained in this work prove the feasibility of a system for storing in superconducting rings energy resulting from mechanical work and its subsequent reversibility (into its former state). This procedure has two advantages over the SMES of superconducting coils: the magnetic flux in the rings remains constant during the whole process and, as a result, there is no decrease in critical current. The only forces found in the operations of loading and unloading would be those of compression. The energy stored by the two rings in question is represented by the area under the curve and the x axis ($\int F dx$).
In Ec.8 (repulsion) we note that the values of *I become* very large when the mutual inductance M approaches the values of L (self-induction).(It can be seen that the repulsion curves can store much more energy that the attraction curves would).

**Acknowledgments**
This work is a part of Research Project 09REM007383PR. The authors are indebted to the Consellería de Economia e Industria (Xunta de Galicia) for financial support.

We also want to express our gratitude to the student Xavier Garcia Casas for his assistance in the process of taking accurately the experimental measurements.[1] H.González-Jorge, J.Peleteiro, E.Carballo, L.Romaní, and G.Domarco. Procedure to induce a persistent current in superconducting cylinders or rings. *App.Phys.Lett*, 81(22):4207–4208, 2002.

[2] I.Karaca. Characterization of a cylindrical superconductor disk prepared by the wet technique with microstructure analysis and levitation force measurements using a permanent magnet. *Chinese Journal of Physics*, 47(5):690–696, October 2009.

[3] A.Sanchez, N. Del Valle, E.Pardo, Chen, Du-Xin and C.Navan; Magnetic levitation of superconducting bars, *J.App.Phys*. Vol.99,pp113904 (2006)..

[4] E.Durand. *Magnetostatique*.Masson et Cie,120,Boul, Saint Germain, Paris

[5] Hans-Gerd Küschner, Bertrand Lücke, Helmut Piel, BeateLehndorf. Levitation-force measurements and field mapping of melt-processed YBaCuO ceramics. *Elsevier ,Physica C*,page 247, January, 2000.

[6] Y.D.Chen, Comparison of magnetic levitation force between a permanent magnet and a high temperature superconductor using different force calculation methods. *Physica C*, Vol.372-376 pp1491-1494 (2002).

[7] E.H.Brandt, Superconducting Disks and Cylinders in axial magnetic field, *Physical Review B,*Vol 58,pp.6506-6528 (1998).

[8] N.Yamaki, T.Nishikawa, N.Sakai, K.Sawa, M.Murakami, Levitation forces of bulk superconductors in varying fields, *Physica C*, 392-396 (2003),